\documentclass{article}

\usepackage{arxiv}

\usepackage[utf8]{inputenc} 
\usepackage[T1]{fontenc}    
\usepackage{hyperref}       
\usepackage{url}            
\usepackage{booktabs}       
\usepackage{amsfonts}       
\usepackage{nicefrac}       
\usepackage{microtype}      
\usepackage{lipsum}
\usepackage{graphicx}
\usepackage{amsmath}
\graphicspath{ {./images/} }

\usepackage{algpseudocode}
\usepackage{algorithm}

\title{FPGA-based Neural Network Accelerators for Millimeter-Wave Radio-over-Fiber Systems}

\author{
 Jeonghun Lee \\
  School of Engineering\\
  Royal Melbourne Institute of Technology\\
  Melbourne, VIC 3000 \\
  \texttt{s3638815@student.rmit.edu.au} \\
   \And
 Jiayuan He \\
  School of Information and Computer Science\\
  The University of Melbourne\\
  Melbourne, VIC 3010 \\
  \texttt{hjhe@student.unimelb.edu.au} \\
  \And
 Ke Wang \\
  School of Engineering\\
  Royal Melbourne Institute of Technology\\
  Melbourne, VIC 3000 \\
  \texttt{ke.wang@rmit.edu.au} \\
}

\begin{document}
\maketitle
\begin{abstract}
With the rapidly-developing high-speed wireless communications, the 60 GHz millimeter-wave (mm-wave) frequency range has attracted extensive interests, and radio-over-fiber (RoF) systems have been widely investigated as a promising solution to deliver mm-wave signals. Neural networks have been proposed and studied to improve the mm-wave RoF system performances at the receiver side by suppressing both linear and nonlinear impairments. However, previous studies of neural networks in mm-wave RoF systems all focus on the use of off-line processing with high-end GPUs or CPUs, which are not practical for low power-consumption, low-cost and limited computation platform applications. To solve this issue, in this paper we investigate neural network hardware accelerator implementations for mm-wave RoF systems for the first time using the field programmable gate array (FPGA), taking advantage of the low power consumption, parallel computation, and reconfigurablity features of FPGA. Both convolutional neural network (CNN) and binary convolutional neural network (BCNN) hardware accelerators are demonstrated. In addition, to satisfy the low-latency requirement in mm-wave RoF systems and to enable the use of low-cost compact FPGA devices, a novel inner parallel computation optimization method for implementing CNN and BCNN on FPGA is proposed. It is shown that compared with the popular embedded processor (ARM Cortex A9) execution latency, the proposed FPGA-based hardware accelerator reduces the processing delay in mm-wave RoF systems by about 99.45\% and 92.79\% for CNN and BCNN, respectively. Compared with non-optimized FPGA implementations, results show that the proposed inner parallel computation method reduces the processing latency by about 44.93\% and 45.85\% for CNN and BCNN, respectively. In addition, compared with the GPU implementation, the latency of CNN implementation with the proposed optimization method is reduced by 85.49\%, while the power consumption is reduced by 86.91\%. Although the latency of BCNN implementation with the proposed optimization method is larger compared with the GPU implementation, the power consumption is reduced by 86.14\%. The demonstrated FPGA-based neural network hardware accelerators provide a promising solution for mm-wave RoF systems.
\end{abstract}

\section{Introduction}
High-speed wireless communications are highly demanded by end-users to support various broadband applications, and the 5G technology has been widely studied recently to satisfy the rapidly-growing demand. Due to the high-speed requirements, the use of millimeter-wave (mm-wave) frequency range has attracted intensive interest. However, the mm-wave has high free-space propagation loss and typically requires line-of-sight (LOS) links. To solve these limitations, the millimeter-wave radio-over-fiber (mm-wave RoF) systems have been widely studied by leveraging the advantages of optical fibers, such as the low transmission loss and the broad bandwidth. However, the signals in mm-wave RoF systems are impaired and distorted by a number of linear and nonlinear effects induced during signal modulation, amplification, transmission and detection \cite{Impair1}, such as the fiber chromatic dispersion and nonlinearities, phase noises, optical and electrical amplifications, and the square-law detection of photo-detectors \cite{Laser_nonlinearity,Koonen08}. To overcome these limitations, various techniques have been studied, including both analog processing and digital signal processing techniques \cite{mmwave_dsp1,mmwave_dsp2,DSP_impair,analog_tech}.  However, nonlinear effects are difficult to be solved using these signal processing schemes, and each processing step typically only targets at suppressing one or few impairments, resulting in relatively limited capability \cite{DSP_impair}.

To solve these limitations, neural networks have been proposed and studied to improve the performance of mm-wave RoF systems \cite{ML_ex_mmWaveRoF}. Neural networks are widely considered as equalizers or classifiers \cite{ML_ex1,ML_ex2_RNN,ML_ex3}, and compared with conventional signal processing methods, they are capable of compensating various linear and nonlinear impairments simultaneously. Therefore, better performance has been achieved in mm-wave RoF systems with neural networks \cite{ML_ex4}.

Although neural networks, such as the fully-connected neural network (FC-NN), the convolutional neural network (CNN), and the binary convolutional neural network (BCNN), have shown promising capabilities in improving the performance of mm-wave RoF systems, they have high computation cost and power consumption \cite{CNN_workload,FPGAbeatsGPU}. Previous studies have mainly focused on the development and adaptation of neural networks in optical communication systems, and GPUs or CPUs in high performance computers have been used to implement neural networks \cite{OptCom_gpu_ex1,OptCom_gpu_ex2}. However, the use of such high profile platforms is not practical in many real applications, such as in base stations, due to the high cost and high power consumption \cite{FPGAbeatsGPU}. In addition, the latency is a critical issue in mm-wave RoF systems, and the latency requirement will be even more stringent in future wireless communications. However, previous studies have not considered the additional latency of neural networks in mm-wave RoF systems. Thus, power efficient, low-latency, low-cost and practical hardware implementations of neural network signal processors (i.e., neural network hardware accelerators) in mm-wave RoF systems are highly demanded.

Neural network hardware accelerators have been studied in several applications, such as the image recognitions \cite{CNN_ImageRecog}. Both application-specific integrated circuits (ASIC) chips and field programmable gate arrays (FPGAs) have been studied for neural network hardware accelerators. ASIC chips have the advantages of high speed and low power consumption. However, the cost of designing ASIC chips is high, and ASIC chips typically only target at one or few specific tasks, limiting its fabrication and application in large volume with flexibility. On the other hand, FPGA devices are capable of implementing flexible and reconfigurable neural network algorithms, and hence, they have the potential to be used in various applications flexibly. In addition, the FPGA-based neural network accelerator also has parallel computation capabilities with low power consumptions \cite{FPGAbeatsGPU,FPGAbenefit}. Therefore, the FPGA has been considered as a promising candidate for neural network accelerator designs.

 Previous FPGA-based neural network hardware accelerators are mainly used for relatively large datasets, which require considerably high resource usages that lead to proportional processing latency. To solve the long processing delay limitation, several implementation optimization methods for FPGAs to accelerate the neural network computation have been studied \cite{CNN_workload}. Unrolling loops for convolution computations in CNNs (and BCNNs) \cite{unrolled_opt} and pipelining their computations have been widely adopted in image recognition applications to reduce the latency \cite{pipeline_opt}. However, it requires considerably large amount of hardware resources in FPGA devices to maximize parallel computations and it also leads to very high power consumption. Therefore, this method is not practical for applications requiring low power consumption, small amount of hardware resources (i.e., low cost) and low processing latency, such as in mm-wave RoF systems, especially at the base stations or remote access points.
 
 In this paper, to the best of our knowledge, we propose and demonstrate FPGA-based neural network hardware accelerators for mm-wave RoF system applications for the first time. Both CNN and BCNN FPGA hardware accelerators are investigated experimentally. To achieve the low latency, low power consumption and low cost implementation requirements, a novel inner parallel hardware optimization method is also proposed. With the proposed optimization method, the neural network hardware accelerator can be implemented using compact FPGA platforms with limited hardware resources and low power consumption. Experimental results show that the signal processing latency is reduced by about 44.93\% and 45.85\% for CNN and BCNN, compared with the FPGA-based neural network accelerators without the optimization method. Compared with the popular ARM Cortex A9 embedded processor, the proposed optimization method achieves latency reductions of 99.45\% and 92.79\% for CNN and BCNN, respectively. Results also show that the BER performance of the mm-wave RoF system with proposed FPGA-based hardware accelerators is similar with that of the system with neural networks implemented using high-end GPUs, whilst the cost and power consumption are significantly reduced. For CNN and BCNN, the realized power reductions are 86.91\% and 86.14\%, respectively. The latency of CNN FPGA-based hardware accelerator is also substantially reduced by 85.49\% compared with the GPU case. Therefore, the demonstrated FPGA-based CNN and BCNN hardware accelerators provide a promising solution for practical applications in mm-wave RoF systems. The major contributions of this paper are summarized as follows:
\begin{itemize}
  \item The FPGA-based CNN and BCNN hardware accelerators have been proposed and studied for mm-wave RoF systems for the first time.
  \item The proposed CNN and BCNN FPGA hardware accelerators have been experimentally demonstrated in a 60 GHz RoF system. Compared with the results obtained using neural networks implemented by GPUs, similar BER performance has been achieved, whilst the power consumption is significantly reduced. Much lower latency is also achieved with the CNN FPGA hardware accelerator.
  \item A novel optimization method based on inner parallel computation has been proposed, enabling lower latency and the implementation of CNN using a compact FPGA platform to achieve lower power consumption and lower cost. The hardware friendly Leaky-ReLU function \cite{HW_leakyrelu} has been utilized to further reduce the number of logic resources required and the power consumption. With the proposed optimization method, the neural network hardware accelerator has been realized using low-cost compact FPGAs without any BER performance degradation.
\end{itemize}

\section{FPGA-based CNN and BCNN Hardware Accelerators for Mm-wave RoF Systems}

\begin{figure}[h]
\centering
\fbox{\includegraphics[width=80mm]{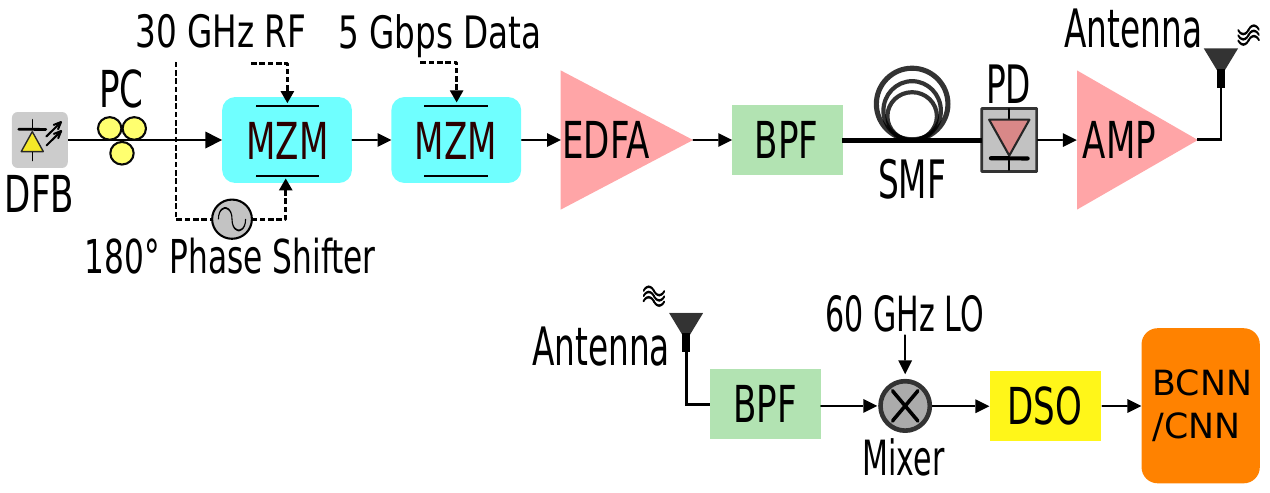}}
\caption{The architecture of 60 GHz mm-Wave RoF system with CNN- and BCNN-based decision schemes.  DFB: distributed feedback laser; PC: polarization controller; MZM: Mach Zehnder modulation; EDFA: Erbium doped fiber amplifiers; BPF: bandpass filter; SMF: single-mode fiber; PD: photo-detector; AMP: amplifiers; LO: local oscillators; and DSO: digital sampling oscilloscope.}
\label{fig:RoF_System_Exper}
\end{figure}

\subsection{Mm-Wave RoF System with CNN and BCNN Decision Schemes}

Neural network based decision schemes can be used to suppress various impairments and to achieve improved BER performance in mm-wave RoF systems \cite{ML_ex4, RoF_CNN_BCNN}. Compared with the FC-NN scheme, it has been shown that the CNN and BCNN schemes can achieve slightly better BER performance with reduced computation cost \cite{RoF_CNN_BCNN}. Therefore, in this paper we study the FPGA-based neural network hardware accelerators based on CNN and BCNN architectures. The general system architecture of the mm-wave RoF system considered is shown in Figure \ref{fig:RoF_System_Exper}, where the double-sideband carrier-suppression (DSB-CS) modulation scheme is utilized. The CNN or BCNN based decision scheme is implemented at the receiver side. As discussed in the previous section, the neural networks in previous studies have been implemented using high-end GPUs, which is not practical in RoF system applications due to the cost and power consumption considerations. In addition, the latency of neural networks, which is critical for wireless communications, has not been considered in previous studies.
 
\begin{figure}[h]
\centering
\fbox{\includegraphics[width=8cm, height=6cm]{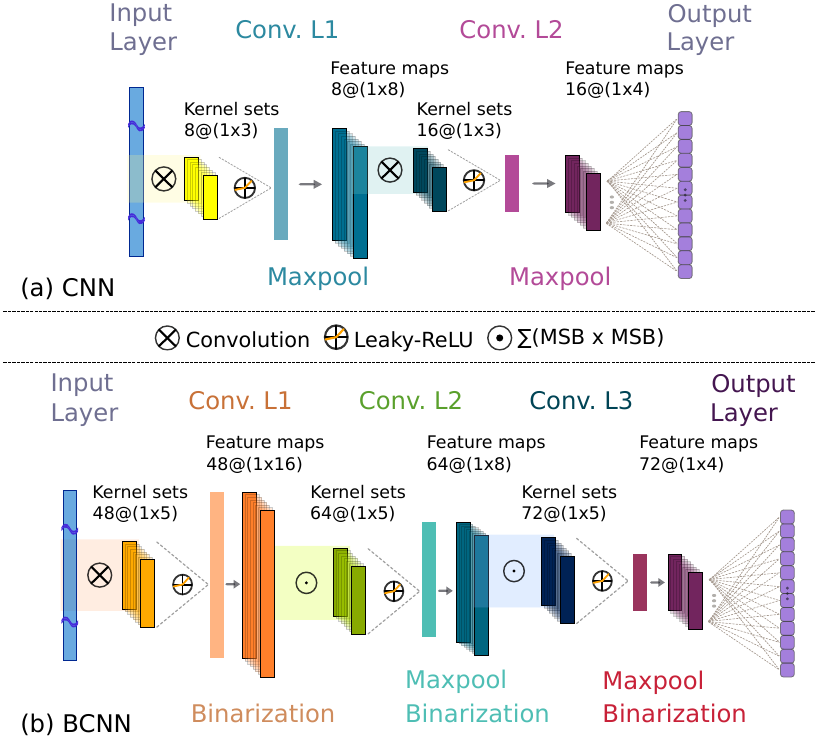}}
\caption{The architecture of neural network decision schemes for mm-wave RoF systems. (a) CNN; and (b) BCNN.}
\label{fig:RoF_CNN_BCNN_Archit}
\end{figure}

To facilitate the description and discussion of proposed FPGA-based neural network hardware accelerators, here we briefly show the working principles and architectures of the CNN and BCNN based decision schemes. More details are available in \cite{RoF_CNN_BCNN}. The CNN decision scheme is shown in Figure \ref{fig:RoF_CNN_BCNN_Archit}(a), which consists of the input layer, 2 convolutional layers, and the output layer. Our CNN decision scheme carries out 1-D convolution operation, followed by 1-D maxpooling and Leaky-ReLU nonlinear activation function. The output layer is a fully-connected layer that computes multiplications and additions to generate the final symbol decision. The 1-D convolution computation in the CNN decision scheme can be expressed as \cite{RoF_CNN_BCNN}
\begin{equation}
\emph{Conv}^{\emph{n}} = \displaystyle \emph{B}^{\emph{n}} + \sum_{1}^{\emph{N}}\sum_{1}^{\emph{R}}\sum_{1}^{\emph{F}}\emph{X}^{\emph{n}} \otimes \emph{K}^{\emph{n}} 
\label{eq_cnn}
\end{equation}
where \emph{B} is the bias parameter, \emph{X} is the input data of each convolutional layer, \emph{K} is the kernel set, \emph{Conv} is the outcome of the convolution computation, \emph{n} is the layer number, \emph{N} is the number of kernel sets, \emph{R} is the data size , and \emph{F} is the kernel size. By conducting the convolution computation, signal characteristics carried by received symbols can be effectively learned, and the impairments and distortions can be suppressed, such as the inter-symbol interference (ISI). 

The BCNN based decision scheme is shown in Figure \ref{fig:RoF_CNN_BCNN_Archit}(b), which consists of the input layer, 3 convolutional layers, and the fully-connected output layer. The major difference from CNN is that the convolution computation is implemented by using the most-significant-bit (MSB), i.e., the sign bit, multiplications and additions, which can be expressed as \cite{RoF_CNN_BCNN}
\begin{equation}
\emph{Binary-Conv}^{\emph{n}}
= \displaystyle \emph{B}^{\emph{n}} + \sum_{1}^{\emph{N}}\sum_{1}^{\emph{R}}\sum_{1}^{\emph{F}}\emph{MSB}\left(\emph{X}^{\emph{n}}\right) \times \emph{MSB}\left(\emph{K}^{\emph{n}}\right)
\label{eq_bcnn}
\end{equation}
where the notations for variables are the same as the CNN case above. 

For both CNN and BCNN based decision schemes, in their convolutional layers, the Leaky-ReLU is used as the nonlinear activation function, which can be expressed as:
 \begin{equation}
 \label{eq_leaky_relu}
  \emph{f(x)}=\begin{cases}
    0.25\cdot \emph{x}, & \text{$\emph{x}<0$}\\
    \emph{x}, & \text{$\emph{x} \geq 0$}
  \end{cases}
\end{equation}

The leaky-ReLU activation function is selected in the CNN and BCNN decision schemes since it can achieve the best accuracy and avoid the gradient vanishing problem during the training process \cite{grad_vanishing}. It is also selected due to the potential of reducing the need for hardware resources, as will be discussed later in Section 2.3.

\subsection{FPGA-based Neural Network Hardware Accelerator for Mm-wave RoF Systems}

Although neural networks have shown great capabilities in solving the limitations in mm-wave RoF systems, they require considerably high computation cost and induce relatively long latency. In addition, high performance computing platforms equipped with GPUs are widely used to implement neural networks, which also results in high power consumptions and high costs. To solve these limitations, here we propose and study the FPGA-based CNN and BCNN hardware accelerators for the mm-wave RoF system. We focus on CNN and BCNN hardware accelerators here, since they have shown better capability and lower computation cost than the FC-NN in RoF systems in our previous study \cite{RoF_CNN_BCNN}.


\begin{figure}[h]
\centering
\fbox{\includegraphics[width=8cm, height=4cm]{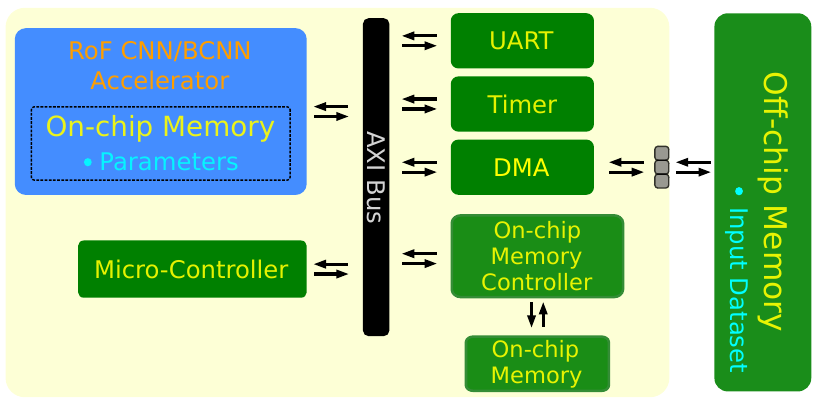}}
\caption{The overall architecture of FPGA-based CNN or BCNN hardware accelerator for mm-wave RoF systems.}
\label{fig:FPAG_Archit}
\end{figure}

The overall architecture of FPGA-based CNN and BCNN hardware accelerators for mm-wave RoF systems is depicted in Figure \ref{fig:FPAG_Archit}. The neural network hardware accelerator mainly consists of the micro-controller, direct memory access (DMA), off-chip memory, on-chip memory controller, on-chip memory, CNN/BCNN decision scheme IP for RoF systems, and AXI bus system. The on-chip memory, which is typically the distributed block RAM (BRAM), is used to store the weight and bias parameters of the neural network decision scheme, and the off-chip memory is used to store the received signal (i.e., dataset) of the mm-wave RoF system. The received signal stored in the external memory is transferred to the implemented CNN- and BCNN-based decision schemes via the AXI bus system controlled by the DMA, which is an IP for controlling data flows from the external memory to the on-chip memory. Then the received signal of the mm-wave RoF system is processed with the weight and bias parameters stored in the on-chip memory during the inference period. The sequence and flow of data and neural network parameters are controlled by the soft-IP micro-controller, which can be programmed with C or other FPGA Software Development Kit (SDK) tools. We use the Timer and the universal asynchronous receiver transmitter (UART) blocks to measure the latency and to show the decision results from the CNN or BCNN decision scheme, respectively. 

\begin{figure}[h]
\centering
\fbox{\includegraphics[width=13cm, height=10cm]{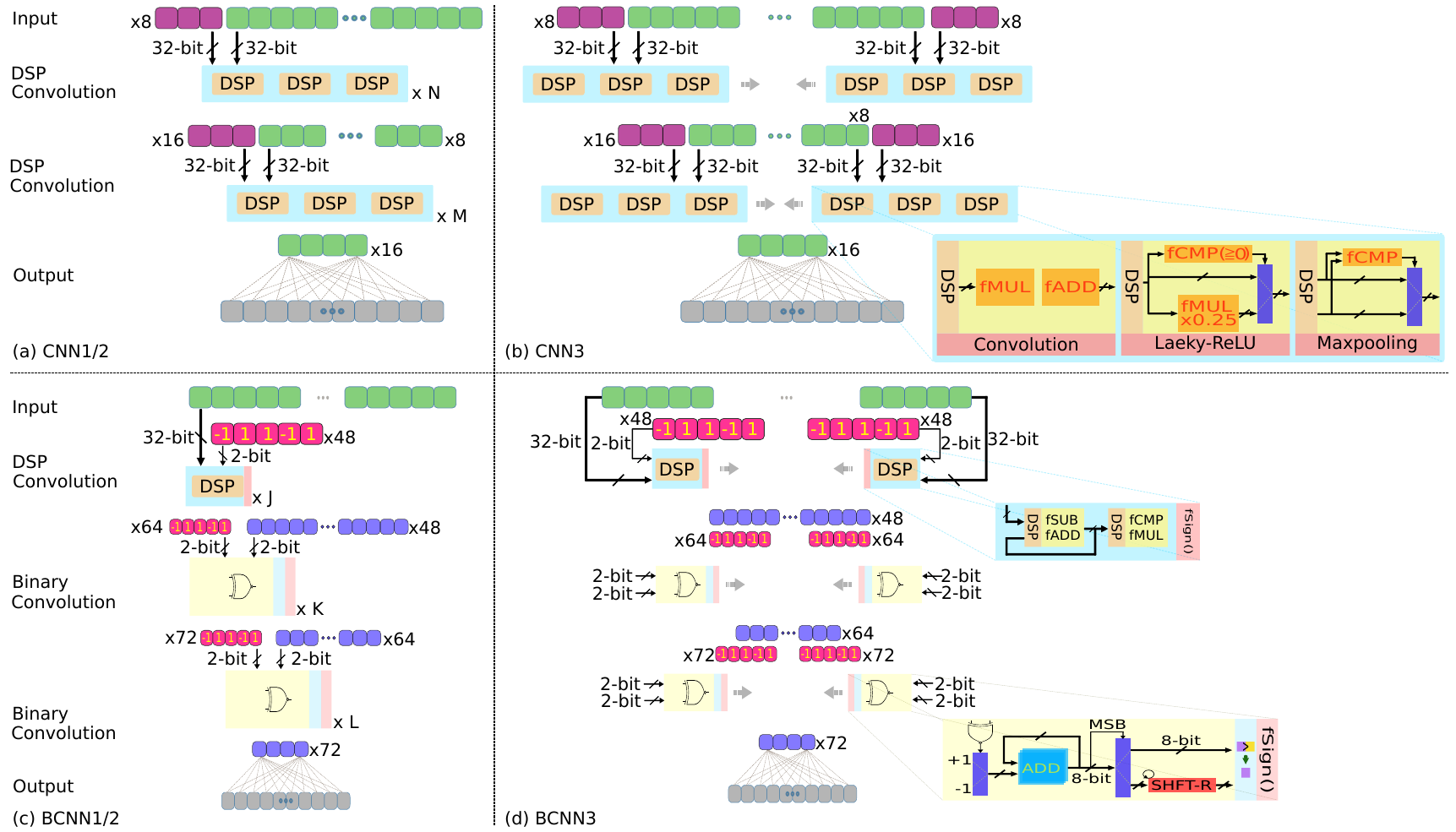}}
\caption{The architecture of FPGA-based CNN and BCNN hardware accelerators. (a) CNN1 and CNN2 architectures; (b) CNN3 architecture with the proposed optimization method; (c) BCNN1 and BCNN2 architectures; and (d) BCNN3 architecture with the proposed optimization method.}
\label{fig:FPGA_RoF_Archit}
\end{figure}

As discussed in the previous section, low latency, low hardware resource requirements and low power consumption are highly desired in the FPGA-based neural network accelerators for mm-wave RoF systems. To realize these requirements, the CNN and BCNN based decision schemes are implemented in 3 different hardware accelerator architectures, and they are shown in Figure \ref{fig:FPGA_RoF_Archit}. Figure \ref{fig:FPGA_RoF_Archit}(a) depicts the CNN1 and CNN2, which are implemented with the non-optimized method and the fully unrolled and pipelining optimization method (widely used in image recognition applications) \cite{unrolled_opt,pipeline_opt}, respectively. In both CNN1 and CNN2 architectures, the convolutional layers are implemented sequentially. Thus, the multiplications and additions are followed by the maxpooling and Leaky-ReLU function operations in each convolutional layer, before the operations in the next convolutional layer can be executed. After all convolutional layers, the multiplication and addition operations in the fully-connected output layer are then executed to generate the final symbol decisions. The notations \emph{N} and \emph{M} in Figure \ref{fig:FPGA_RoF_Archit}(a) represent the number of parallel computation units in the convolutional layers. For CNN1, both \emph{N} and \emph{M} equal to 1 and for CNN2, they are larger than 1 and are determined by the synthesis strategy and the available resources. Comparing the CNN1 and CNN2 architectures, it can be seen that the CNN2 uses parallel computations, and hence, it is capable of reducing the processing delay caused by the neural network, which is important for mm-wave RoF applications. However, as will be discussed later, the FPGA implementation of CNN2 requires a significantly larger amount of hardware resources and leads to higher power consumptions, which set practical application limitations. 

Figure \ref{fig:FPGA_RoF_Archit}(c) illustrates the BCNN1 and BCNN2 architectures, which are similarly implemented with the non-optimized method and the fully-unrolled and pipelining optimization method \cite{unrolled_opt,pipeline_opt}, respectively. Similar with the CNN case, BCNN1 and BCNN2 also execute convolutional layers sequentially, and the major difference is that the convolutional layers are binarized. The notations of \emph{J} / \emph{K} / \emph{L} also represent the number of parallel computation units, which are set to 1 in BCNN1 and decided by the synthesis strategy and the available resources of FPGA platforms in BCNN2. Similar with the CNN hardware accelerator case, the BCNN2 architecture uses parallel computation to reduce the processing latency, at the cost of higher power consumptions and larger amount of hardware resources.

Due to the different data types used during computations, the dominant computation units required in the CNN and BCNN hardware accelerators are different. In FPGA-based CNN hardware accelerators, 32-bit floating point numbers in the IEEE-754 standard \cite{IEEE754_CNN} are used for computations, and hence, as shown in Figure \ref{fig:FPGA_RoF_Archit}(a), the DSP IP blocks are used for their hardware implementations. More specifically, the multiplications (fMUL) and additions (fADD) in the convolution computations described by Eq.\ref{eq_cnn} are implemented with DSP IP blocks. The DSP IP blocks are also needed in the following 1-D maxpooling and Leaky-ReLU operations in the convolutional layers, since they require comparison (fCMP) and multiplication (fMUL) computations, which also use 32-bit floating point numbers. The fully-connected output layer requires DSP IP blocks as well, since the output layer also requires multiplications and additions using 32-bit floating point numbers. Therefore, a large number of DSP IP blocks are required in the FPGA-based CNN hardware accelerators.

On the other hand, in the BCNN hardware accelerators, binary convolutional layers are mainly used, except the first layer (the first layer processes the received data). The binary numbers in FPGAs are normally handled with logic gates instead of DSP IP blocks. Therefore, in the BCNN hardware accelerators shown in Figure \ref{fig:FPGA_RoF_Archit}(c), the first convolutional layer is implemented with DSP IP blocks, since 32-bit floating point numbers (i.e., the input data) are processed in this layer to achieve high accuracy. The following convolutional layers, which execute the binary convolution computation using the most significant bit (MSB) as described in Eq.\ref{eq_bcnn}, are realised with XNOR logic gates. In the output layer, since it also processes binary numbers, the logic gates are utilized. Therefore, the logic gates in FPGAs are mainly used in the BCNN hardware accelerator implementations.

In addition, as shown by Eq.\ref{eq_cnn} and Eq.\ref{eq_bcnn}, the convolution and binary convolution operations require a large amount of multiplications, additions and subtractions (for BCNN only). Therefore, when the fully unrolled and pipelining optimization method is adopted to reduce the processing latency, considerably high amounts of hardware resources are required. Specifically, a very large number of DSP IP blocks are needed for the CNN2 architecture, and a very large amount of logic gates are required for the BCNN2 architecture. The hardware requirements and usages also result in high power consumptions in the CNN2 and BCNN2 FPGA implementations.

\subsection{Inner Parallel Computation Optimization for FPGA-based CNN and BCNN Hardware Accelerators in mm-wave RoF Systems}

\begin{algorithm*}
\caption{Inner parallel optimization method for convolution computation}\label{alg:inner parallel convolution}
\begin{algorithmic}
\Procedure{Convolution Operation}{$Data,Weights$}
\For{$ m=0 ; m < M ; m++ $} \Comment{Output channel loop}
\For{$ n=0 ; n < N ; n++ $} \Comment{Input channel loop}
\For{$ i=0 ; i <I/2 ; i++ $} \Comment{Feature map loop}
\For{$ k=0 ; k < K ; k++ $} \Comment{Kernel loop}
\State $ Conv[m][n][\quad\;\;\,i\quad\;\;\,]+= X[n][S*\qquad i\qquad\;+ k] \otimes Kernel[m][n][k] $ 
\State $ Conv[m][n][I-i-1 ]+= \displaystyle X[n][S*(I-i-1)+k] \otimes Kernel[m][n][k] $ 
\If{k==(K-1)} \Comment{Activation}
\State $ out-relu[m][\quad\;\;\, i\quad\;\;\,]+= Leaky(Conv[m][\quad\;\;\,i \quad\;\;\,]) $ 
\State $ out-relu[m][I-i-1]+= \displaystyle Leaky(Conv[m][I-i-1]) $ 
\EndIf
\If{(i\%2)==1} \Comment{Maxpooling}
\State $ out-mxpl[m][\qquad i/2\qquad\;]+= MP(Conv[m][\qquad i/2\qquad\,]) $ 
\State $ out-mxpl[m][(I-i-1)/2]+= MP(Conv[m][(I-i-1)/2]) $ 
\EndIf
\EndFor\label{euclidendfor}
\EndFor\label{euclidendfor}
\EndFor\label{euclidendfor}
\EndFor\label{euclidendfor}
\EndProcedure
\end{algorithmic}
\end{algorithm*}

As discussed in the previous section, the un-optimized FPGA-based CNN and BCNN hardware accelerators do not support parallel computation, which results in relatively long latency, whilst the ones with the fully unrolled and pipelining optimization method require a large amount of hardware resources and have high power consumptions. To solve these limitations and to meet the requirements of mm-wave RoF systems, which demand low latency, low computation hardware requirement and low power consumption, here we propose the inner parallel computation optimization method and the use of hardware friendly Leaky-ReLU nonlinear activation function \cite{HW_leakyrelu} for the FPGA-based CNN and BCNN hardware accelerators. We refer them as CNN3 and BCNN3, respectively, and their architectures are illustrated in Figure \ref{fig:FPGA_RoF_Archit}(b) and Figure \ref{fig:FPGA_RoF_Archit}(d), respectively. The proposed optimization method is described in detail in Algorithm \ref{alg:inner parallel convolution}. In the proposed optimization method, the convolution operation, which consists of multiplications (fMUL), additions (fADD) and subtractions (fSUB), is computed in parallel from both the start and the end sides of the input data stream, which is stored in the on-chip memory with known addresses. After the convolution computation, the nonlinear activation functions and the maxpooling are also computed in parallel. Because the proposed method computes in parallel from both sides of data stored in on-chip memories, the processing latency can be improved compared with the non-optimized implementations (i.e., CNN1 and BCNN1). In addition, compared with the CNN2 and BCNN2, the proposed optimization method requires less hardware resources and has lower power consumption.

The hardware resources required in the proposed CNN3 and BCNN3 hardware accelerators can be further reduced by using the Leaky-ReLU nonlinear function. This is because that the Leaky-ReLU function can be realized with the arithmetic right shift operation, which can be synthesized with logic gates in FPGAs instead of using DSP IP blocks \cite{HW_leakyrelu}. Compared with the Leaky-ReLU function implemented with DSP IP blocks, the Leaky-ReLU realized with the arithmetic right shift operation requires less hardware resources, and hence, the computation cost and power consumptions are further improved. The use of DSP IP blocks and logic gates in the Leaky-ReLU function can also be reduced by optimizing the co-efficient for the negative results of convolution computation in the function as expressed in Eq.\ref{eq_leaky_relu} \cite{HW_leakyrelu}. We experimentally optimized the co-efficient and it is selected as 0.25 to minimize the use of hardware resources whilst maintaining the BER performance.

\section{Experiments and Results}
\begin{figure}[!ht]
\centering
\fbox{\includegraphics[width=12cm, height=3cm]{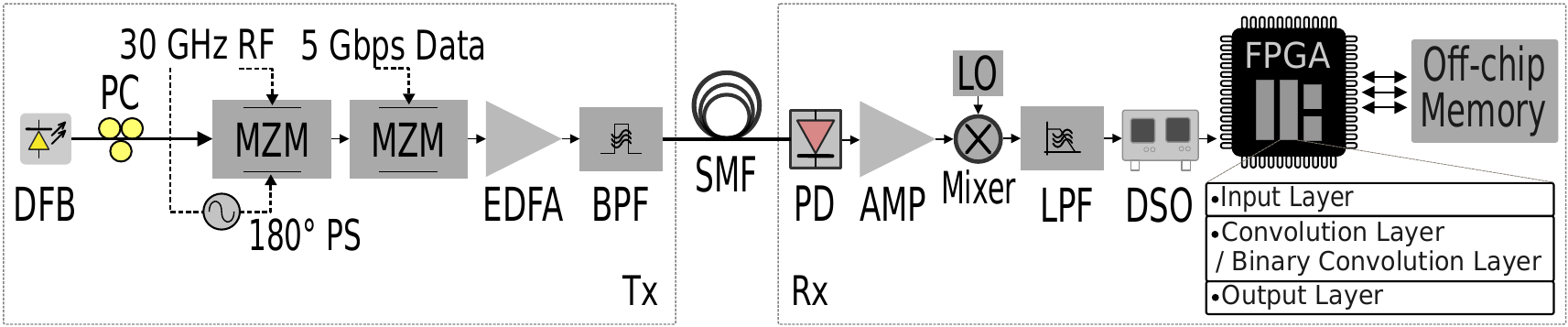}}
\caption{Experimental setup of the FPGA-based CNN/BCNN hardware accelerator for the 60 GHz mm-wave RoF system.}
\label{fig:Exp_FPGA_CNN_BCNN_HW_Acc_RoF}
\end{figure}

\subsection{Experimental Setup}
The FPGA-based CNN and BCNN hardware accelerators for mm-wave RoF systems was experimentally demonstrated using the setup illustrated in Figure \ref{fig:Exp_FPGA_CNN_BCNN_HW_Acc_RoF}. The 60 GHz frequency was used, and the DSB-CS scheme was adopted at the transmitter side, where a DFB laser at 1550 nm served as the light source and a dual-driven Mach-Zehnder modulator (MZM) driven by two complementary 30 GHz RF signals achieved the optical carrier suppression. 5 Gb/s signal to be transmitted was modulated using another MZM. The modulated optical wave was then amplified by an Erbium doped fiber amplifier (EDFA), filtered by an optical bandpass filter (BPF), transmitted via the single-mode fiber (SMF), and detected with a high-speed PIN photo-detector (PD). Due to device limitations, the wireless propagation part was not included in the experiment, and the converted electrical signal was down-converted directly by a RF mixer. To suppress the impairments in the system and to improve the BER performance, the detected signal was then processed by the CNN or BCNN based decision scheme. A high-speed digital sampling oscilloscope (DSO) was used before the CNN or BCNN decision scheme to serve as the analog-to-digital converter (ADC). In the experiment, the CNN and BCNN decision schemes were implemented using both the GPU and the FPGA hardware accelerators proposed and discussed in the previous section.

\begin{table}[h!]
\centering
\caption{\bf FPGA specifications}
\begin{tabular}{ccccc}
\hline
FPGA platform & BRAM & DSP & FF & LUT \\
\hline
Arty-7(XC7A35T) & 50 & 90 & 41,600 & 20,800 \\
VC709(XC7VX690T) & 2940 & 3600  & 866,400 & 433,200 \\
\hline
\end{tabular}
\label{tab:fpga_spec}
\end{table}

The FPGA-based CNN and BCNN hardware accelerators for mm-wave RoF systems were realized using two FPGA platforms, i.e., Xilinx VC709 and Xilinx Arty-7, and their specifications are compared in Table \ref{tab:fpga_spec}, including the number of BRAM, DSP IP blocks, flip-flop (FF) and look-up table (LUT), which are the fundamental reconfigurable resources in the FPGA. It is clear from the table that the Xilinx VC709 FPGA has a significantly larger number of resources available, whilst the Xilinx Arty-7 is more resource-limited. We selected to use these two FPGAs here to show the capability of implementing the proposed neural network hardware accelerates for mm-wave RoF systems on both the high-end and the compact FPGA platforms, which can satisfy different application scenarios, such as in base stations and in embedded devices.

More specifically, the VC709 platform has the Virtex-7 FPGA device with 4GB external DDR3 Synchronous Dynamic Random Access Memory (SDRAM), and the Arty-7 board has the 7-series FPGA device with 256MB DDR3 SDRAM off-chip memory. As discussed in Section 2.2, the off-chip DDR3 SDRAMs were used to store the measured datasets (i.e. received signal of the mm-wave RoF system). Xilinx Vivado High Level Synthesis (HLS, version 2017.3) was used to generate the CNN and BCNN source code programmed with C and to synthesize for the hardware description languages (HDL). The Vivado Design Suite was then used to implement the CNN and BCNN decision schemes for the 60 GHz mm-wave RoF system on the two FPGA platforms. The operating clock speeds for the VC709 platform and the Arty-7 platform were 100 MHz and 83 MHz, respectively.

\subsection{Results and Discussions}

\begin{figure}[!h]
\centering
\fbox{\includegraphics[width=13cm, height=4cm]{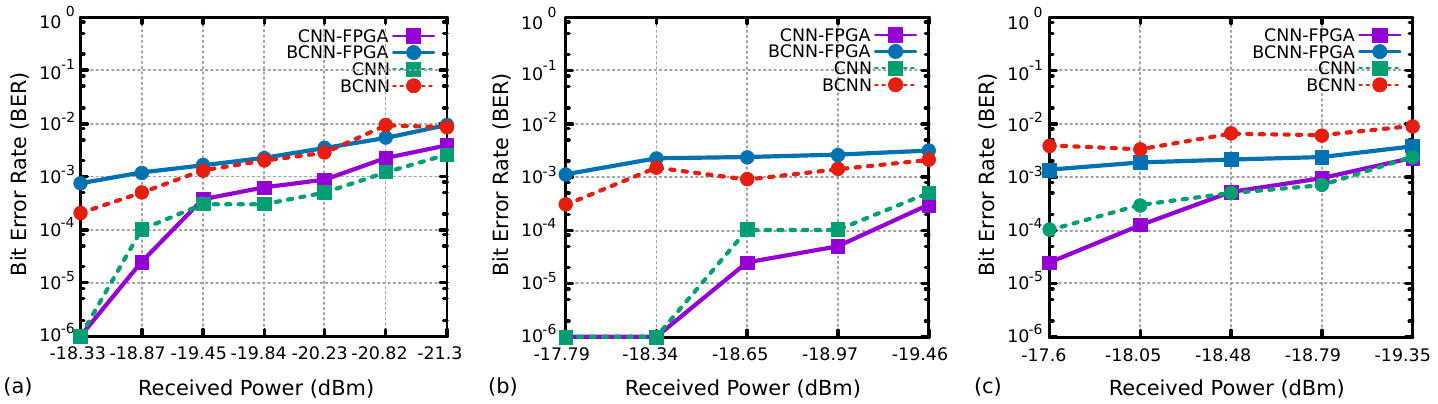}}
\caption{Experimental results on the BER performance of the 60 GHz mm-wave RoF system. (a)fiber length = 10 km; (b)fiber length = 15 km; and (c)fiber length = 20 km.}
\label{fig:RoF_CNN_BCNN_Res}
\end{figure}

To demonstrate the feasibility of proposed FPGA-based CNN and BCNN hardware accelerators in the 60 GHz mm-wave RoF system, we firstly measured the BER performance at different fiber transmission distances. As the comparison benchmark, we also processed the received signal with the CNN and BCNN decision schemes implemented using the GPU, nVidia M5000M. The results are shown in Figure \ref{fig:RoF_CNN_BCNN_Res}. It can be seen that for up to 20 km optical fiber transmission distance, BER performances within the forward-error correction (FEC) limit can be achieved in the mm-wave RoF system with the FPGA-based CNN or BCNN hardware accelerator. It is also clear that similar BER performances are achieved for all tested transmission distances when the neural networks are implemented with either the GPU or the FPGA, confirming the capability of the proposed FPGA-based CNN and BCNN hardware accelerators in RoF systems. All three CNN and three BCNN FPGA-based hardware accelerator architectures as shown in Figure \ref{fig:FPGA_RoF_Archit} were implemented. There is no BER performance difference observed during the measurement, since all hardware architectures implemented the same CNN or BCNN structure as shown in Figure \ref{fig:RoF_CNN_BCNN_Archit}. In addition, compared with the FPGA-based BCNN hardware accelerator, better BER performance can be realized using the CNN implementation for all fiber transmission distances. The worse BER performance of the BCNN hardware accelerator is mainly due to the reduced bit sizes for variables (i.e., binary values) and the additional losses during binarization.

The hardware resource requirement and the processing latency of the FPGA-based CNN hardware accelerators using the three architectures shown in Figure \ref{fig:FPGA_RoF_Archit}(a) and Figure \ref{fig:FPGA_RoF_Archit}(b) were also experimentally analyzed. The results are presented in Table \ref{tab:CNN Resource} and Table \ref{tab:CNN Performance}, respectively. It can be seen that with the VC709 platform, the non-optimized CNN1 implementation requires 15 DSPs, 48 18Kb BRAMs, and 40.1K LUT logic units. To implement the CNN2 architecture with the fully-unrolled and pipelining optimization, 23.6 times more DSP IP blocks and 2.73 times more LUTs are required. The significantly larger number of hardware resources used in the CNN2 implementation enables about 6.95 times faster processing (i.e., the latency is reduced by 85.62\%). On the other hand, 30 DSP IP blocks, 51 18Kb BRAMs, and 43.1K LUTs are needed to implement the CNN3 architecture with the proposed inner parallel computation optimization method. Compared with the CNN1 implementation, two times more DSP IP blocks and 10\% more 18Kb BRAMs and LUTs are needed. The slightly larger number of hardware resources needed in CNN2 realizes 1.81 times faster processing (i.e., the latency is reduced by 44.93\%).

\begin{table*}[!h]
\centering
\caption{\bf Resource utilization of FPGA-based CNN hardware accelerators}
\begin{tabular}{cccccc}
\hline
Neural & Freq. &  DSP & BRAM & Logic  & Device\\
 Network & (MHz) & & (18Kb) & & \\
\hline
CNN1 & 100 & 15 & 48 & 40.1K & VC709 \\
CNN2 & 100 & 355 & 48.5 & 109.5K & VC709 \\
CNN3 & 100 & 30 & 51 & 43.1K & VC709 \\
\hline
CNN1 & 83 & 15 & 48 & 22.6K & Arty-7 \\
CNN3 & 83 & 26 & 44.5 & 26.4K & Arty-7 \\
\hline
\end{tabular}
  \label{tab:CNN Resource}
\end{table*}

\begin{table*}[!h]
\centering
\caption{\bf Performance comparison of FPGA-based CNN hardware accelerators}
\begin{tabular}{cccccc}
\hline
Neural & Freq. & Latency  & Power(W) & Efficiency & Device\\
 Network & (MHz) & (Sec.) & &Index  &\\
\hline
CNN1 & 100 & 606.1$\mu$ & 3.6872 &  - &  VC709 \\
CNN2 & 100 & 87.1$\mu$ & 4.7289 &  3.03 & VC709 \\
CNN3 & 100 & 333.8$\mu$ & 3.7972 &  15.06 & VC709 \\
\hline
CNN1 & 83 & 1,091$\mu$ & 1.5246 &  - & Arty-7 \\
CNN3 & 83 & 432.8$\mu$ & 1.5565 &  28.83 & Arty-7 \\
\hline
CNN & 667 & 60.4m & 2.6531 & 2.31 & Cortex-A9 \\
CNN & 975 & 2.3m & 29 &  0.98 & GPU(nVidia M5000M) \\
\hline
\end{tabular}
  \label{tab:CNN Performance}
\end{table*}

In addition to the implementation on the VC709 platform, which is a high-end FPGA platform, it is also highly desirable to implement the FPGA-based CNN hardware accelerators using more compact and lower-cost FPGA platforms for practical mm-wave RoF system applications. To satisfy this need, we also implemented the hardware accelerators using the compact and resource-limited Arty-7 platform. Due to the large number of hardware resources needed, CNN2 cannot be implemented on the compact FPGA platform, limiting its practical applications. On the other hand, CNN1 and CNN3 can be realized, and the results are also shown in Table \ref{tab:CNN Resource} and Table \ref{tab:CNN Performance}. It is clear from the results that compared to the VC709 implementations, both CNN1 and CNN3 implementations using Arty-7 require a smaller amount of hardware resources. This is mainly due to the difference between DMAs in the VC709 and Arty-7 platforms. Regarding the latency performance, using the Arty-7 platform, the CNN3 implementation with the proposed inner parallel optimization method achieves over 60.32\% latency reduction against the un-optimized CNN1 implementation.

In addition to the hardware resource requirement and the processing latency, the power consumption is also an important parameter for FPGA-based neural network hardware accelerators. Therefore, we also characterized the power consumption performance of the three CNN hardware accelerator architectures experimentally. The results are shown in Table \ref{tab:CNN Performance}. It can be seen that compared with the un-optimized CNN1, although the processing latency is reduced in CNN2, due to the significantly larger number of hardware resources required, the power consumption is increased by more than 28.25\% when implemented using the VC709 platform. On the other hand, the CNN3 architecture with the proposed optimization method only consumes less than 3\% more power compared with the CNN1 baseline architecture, whilst the latency is reduced by about 44.92\%. In the Arty-7 compact platform, compared with the CNN1, the implementation of CNN3 achieves about 60.32\% latency reduction, at the cost of less than 2\% increase in the power consumption. Therefore, the FPGA-based CNN3 hardware accelerator with the proposed inner parallel optimization method is capable of being implemented on compact and low-cost platforms and achieving significantly reduced latency, with only slightly increased power consumption.

\begin{table*}[!h]
\centering
\caption{\bf Resource utilization of FPGA-based BCNN hardware accelerators}
\begin{tabular}{cccccc}
\hline
Neural & Freq. & DSP & BRAM & Logic  & Device\\
Network & (MHz) &      & (18Kb) &      &  \\
\hline
BCNN1 & 100 &  5 & 84.5 & 28.8K &  VC709 \\
BCNN2 & 100 & 19 & 156 & 113.8K &  VC709 \\
BCNN3 & 100 &  5 & 86 & 40K & VC709 \\
\hline
\end{tabular}
  \label{tab:BCNN Resource}
\end{table*}

\begin{table*}[!h]
\centering
\caption{\bf Performance comparison of FPGA-based BCNN hardware accelerators}
\begin{tabular}{cccccc}
\hline
Neural & Freq. & Latency  & Power(W) &  Efficiency & Device\\
Network & (MHz) & (Sec.)   &          &       Index    & \\
\hline
BCNN1 & 100 & 18.08m & 3.7114 &  - & VC709 \\
BCNN2 & 100 & 1.95m & 5.6331 &  1.72 & VC709 \\
BCNN3 & 100 & 9.79m & 3.7422 &  55.25 & VC709 \\
\hline
BCNN  & 667 & 135.8m & 2.6476 & 2.24   & Cortex-A9 \\
BCNN  & 975 & 2.5m & 27 & 3.39  & GPU(nVidia M5000M) \\
\hline
\end{tabular}
  \label{tab:BCNN Performance}
\end{table*}

In addition to the CNN, the FPGA-based BCNN hardware accelerators were also demonstrated using the three architectures shown in Figure \ref{fig:FPGA_RoF_Archit}(c) and Figure \ref{fig:FPGA_RoF_Archit}(d) with the VC709 platform. The experimental results are shown in Table \ref{tab:BCNN Resource} and Table \ref{tab:BCNN Performance}. The non-optimized BCNN1 as a comparison baseline requires 5 DSP IP blocks, 84.5 18Kb BRAM blocks, and 28.8K LUTs. Similar with the CNN case, a significantly larger number of hardware resources are required to implement the BCNN architecture with the fully unrolled and pipelining optimization (i.e., BCNN2), whilst the implementation of BCNN with the proposed inner parallel optimization method (i.e., BCNN3) only requires slightly more hardware resources. The processing latency and power consumption of the three BCNN FPGA hardware accelerators were also measured. From the results shown in Table \ref{tab:BCNN Performance}, it is clear that although better latency is achieved, the BCNN2 also consumes much higher power (51.78\% higher) compared with the un-optimized BCNN1, in addition to the larger number of hardware resources required. On the other hand, the BCNN3 with the proposed optimization method reduces the latency by about 45.85\% over the BCNN1, whilst the increase on power consumption is negligible (less than 1\%).

Comparing the FPGA-based CNN and BCNN hardware accelerators, it can be seen that the CNN hardware accelerators mainly require DSP IP blocks to support floating point number computations, whilst the BCNN hardware accelerators mostly need BRAMs and LUTs for logic gates. This is consistent with the discussion in Section 2.2. Because of this hardware requirement and due to the limited number of LUTs available in the Arty-7 FPGA platform, currently implementing BCNN1 and BCNN3 in the compact Arty-7 platform is not feasible. However, it is possible to implement BCNN1 and BCNN3 using other compact FPGA platforms with more LUTs, such as the XC7A50T FPGA platform. In addition to the difference on the hardware resource requirement, the FPGA-based CNN and BCNN hardware accelerators also have different power consumption and latency performances. In general, as can be seen from Table \ref{tab:CNN Performance} and Table \ref{tab:BCNN Performance}, the CNN hardware accelerators can achieve more than 1-order-of-magnitude better latency performance with comparable power consumption. Therefore, the CNN hardware accelerators are in general better suited for mm-wave RoF system applications. 

From the results and discussions presented above, it can be concluded that for both CNN and BCNN, compared with the FPGA-based hardware accelerator with architecture 1, the architectures 2 and 3 can achieve improved latency performance at the cost of increased power consumption. To further compare the capability and efficiency  of architectures 2 and 3, here we define the efficiency index parameter, which is the latency improvement at the cost of unit increase in the power consumption, and it can be expressed as:
\begin{equation}
Efficiency\  Index
= \displaystyle \frac{Latency \  Improvement \ Ratio}{Power\  Increase\  Ratio}
\label{eq_lat_power}
\end{equation}
where the latency improvement ratio and the power increase ratio are defined as the corresponding improvement or increase compared with those of the architecture 1 FPGA implementation. Since mm-wave RoF systems require both low latency and low power consumption, the defined efficiency index represents a relatively fair measure on the capability and efficiency of FPGA-based neural network hardware accelerators with different optimization methods. We use the architecture 1 based FPGA implementation as the comparison baseline, since it is the most straight-forward method without further optimization.

The efficiency index are calculated for both CNN and BCNN schemes, and the results are shown in Table \ref{tab:CNN Performance} and Table \ref{tab:BCNN Performance}. It is clear that the FPGA-based hardware accelerators with the architecture 3 achieves better efficiency index compared to the architecture 2. Therefore, the proposed inner parallel optimization method is better suited than the fully unrolled and pipelining method in mm-wave RoF applications. In addition, the proposed optimization method also enables the implementation of CNN hardware accelerator in compact and low-cost FPGA platforms, and hence, it facilitates applications in RoF base stations or embedded devices. Due to the lower power consumption, the efficiency index is even higher when the proposed inner parallel optimization method is implemented using the compact and resource-limited FPGA platform.

The performance of the proposed FPGA-based CNN and BCNN hardware accelerators for mm-wave RoF systems was also compared with one of the popular embedded processors, ARM Cortex A9. The ARM Cortex A9 processor executes instructions and computations in the designed pipeline architecture with a reduced instruction set computer (RISC), and it has been widely used to benchmark the performance and capability of FPGA-based neural network hardware accelerators \cite{CortexA9_FPGA_ex1,Cortex_FPGA_ex2}. The results are also shown in Table \ref{tab:CNN Performance} and Table \ref{tab:BCNN Performance}. It is clear that for both CNN and BCNN based decision schemes, although the clock speed of Cortex A9 is much faster than the clock speeds of VC709 and Arty-7 FPGA platforms, the signal processing latency induced is significantly longer. This is due to the longer floating point number computations and the less parallel architecture in Cortex A9. Specifically, the FPGA-based CNN and BCNN hardware accelerators with the proposed optimization method (i.e., CNN3 and BCNN3) can achieve processing latency reductions of 99.45\% and 92.79\%, respectively, at the cost of about 43.12\% and 41.34\% higher power consumptions.

In addition, the latency and power consumptions of CNN and BCNN GPU implementations were also measured and they are compared in Table \ref{tab:CNN Performance} and Table \ref{tab:BCNN Performance}. Compared with the GPU implementations, the latency of CNN FPGA hardware accelerator with the proposed optimization method is reduced by 85.49\%, together with 86.91\% reduction on the power consumption. For the BCNN, although the latency of the FPGA hardware accelerator with the proposed optimization method is longer compared with the GPU implementation, the power consumption is reduced significantly by 86.14\%.

\section{Conclusions}

In this paper, we have studied and demonstrated FPGA-based CNN and BCNN hardware accelerators for mm-wave RoF systems, and a novel inner parallel computation optimization method has been proposed to further enhance the capabilities of the hardware accelerators. Experimental results have shown that CNN- and BCNN-based decision schemes implemented in FPGA hardware accelerators can achieve similar BER performance as those obtained using GPUs, and the BER within the forward-error-correction (FEC) limit can be achieved for up to 20 km fiber transmission distance.

Three FPGA-based CNN and BCNN hardware accelerators have been implemented and demonstrated. Results have shown that the architecture 1 (i.e., CNN1 and BCNN1) with the non-optimized method requires the smallest number of hardware resources and has the lowest power consumption, whilst the latency is long, which is problematic for RoF applications. The architecture 2 (i.e., CNN2 and BCNN2) improves the latency considerably, whilst the applied optimization method requires a significantly larger amount of hardware resources and has much higher power consumption. The architecture 3 (i.e., CNN3 and BCNN3) with the proposed inner parallel optimization method applied improves the latency substantially, whilst the increase in resources and power consumption is moderate to minimum. To compare the optimization methods, an efficiency index has been defined to measure the capability and efficiency of latency improvement per unit increase of the power consumption. It has been shown that the inner parallel optimization can achieve better efficiency index, and hence, it is better suited for mm-wave RoF applications, which require both low power consumption and low latency simultaneously.

In addition, more general comparisons have been conducted by comparing the performance of FPGA-based CNN and BCNN hardware accelerators with those implemented using the popular embedded processor (ARM Cortex A9) and the GPU (nVidia M5000M). Results have shown that compared with Cortex A9, the FPGA implementations with the proposed optimization method can achieve processing latency reductions of 99.45\% and 92.79\% for CNN and BCNN, respectively, at the cost of moderately increased power consumption (about 43.12\% and 41.34\% for CNN and BCNN). Besides, compared with the GPU implementation, the power consumption of CNN and BCNN FPGA-based hardware accelerators with the proposed inner parallel optimization method is reduced by about 86.91\% and 86.14\%, respectively, and the latency is also reduced by 85.49\% for the CNN case. Therefore, the FPGA-based neural network hardware accelerators demonstrated in this paper provide a promising solution for mm-wave RoF systems.


\bibliographystyle{unsrt}  
\bibliography{references}  

\begin{thebibliography}{10}

\bibitem{Impair1}
Y.~{Tian}, S.~{Song}, K.~{Powell}, K.~{Lee}, C.~{Lim}, A.~{Nirmalathas}, and
  X.~{Yi}.
\newblock A 60-ghz radio-over-fiber fronthaul using integrated microwave
  photonics filters.
\newblock {\em IEEE Photonics Technology Letters}, 29(19):1663--1666, Oct 2017.

\bibitem{Laser_nonlinearity}
D.~{Kedar}, D.~{Grace}, and S.~{Arnon}.
\newblock Laser nonlinearity effects on optical broadband backhaul
  communication links.
\newblock {\em IEEE Transactions on Aerospace and Electronic Systems},
  46(4):1797--1803, Oct 2010.

\bibitem{Koonen08}
A.~M.~J. Koonen and M.~Garc\'{i}a Larrod\'{e}.
\newblock Radio-over-mmf techniques---part ii: Microwave to millimeter-wave
  systems.
\newblock {\em J. Lightwave Technol.}, 26(15):2396--2408, Aug 2008.

\bibitem{mmwave_dsp1}
Z.~{Cao}, J.~{Yu}, M.~{Xia}, Q.~{Tang}, Y.~{Gao}, W.~{Wang}, and L.~{Chen}.
\newblock Reduction of intersubcarrier interference and frequency-selective
  fading in ofdm-rof systems.
\newblock {\em Journal of Lightwave Technology}, 28(16):2423--2429, Aug 2010.

\bibitem{mmwave_dsp2}
Z.~{Cao}, J.~{Yu}, H.~{Zhou}, W.~{Wang}, M.~{Xia}, J.~{Wang}, Q.~{Tang}, and
  L.~{Chen}.
\newblock Wdm-rof-pon architecture for flexible wireless and wire-line layout.
\newblock {\em IEEE/OSA Journal of Optical Communications and Networking},
  2(2):117--121, February 2010.

\bibitem{DSP_impair}
S.~J. {Savory}.
\newblock Digital coherent optical receivers: Algorithms and subsystems.
\newblock {\em IEEE Journal of Selected Topics in Quantum Electronics},
  16(5):1164--1179, Sep. 2010.

\bibitem{analog_tech}
Yiguang Wang, Li~Tao, Xingxing Huang, Jianyang Shi, and Nan Chi.
\newblock Enhanced performance of a high-speed wdm cap64 vlc system employing
  volterra series-based nonlinear equalizer.
\newblock {\em IEEE Photonics Journal}, 7(3):1--7, 2015.

\bibitem{ML_ex_mmWaveRoF}
Yue Cui, Min Zhang, Danshi Wang, Siming Liu, Ze~Li, and Gee-Kung Chang.
\newblock Bit-based support vector machine nonlinear detector for
  millimeter-wave radio-over-fiber mobile fronthaul systems.
\newblock {\em Opt. Express}, 25(21):26186--26197, Oct 2017.

\bibitem{ML_ex1}
Chun-Yen Chuang, Li-Chun Liu, Chia-Chien Wei, Jun-Jie Liu, Lindor Henrickson,
  Wan-Jou Huang, Chih-Lin Wang, Young-Kai Chen, and Jyehong Chen.
\newblock Convolutional neural network based nonlinear classifier for 112-gbps
  high speed optical link.
\newblock {\em Optical Fiber Communication Conference}, page W2A.43, 2018.

\bibitem{ML_ex2_RNN}
Jiayuan He, Jeonghun Lee, Tingting Song, Hongtao Li, Sithamparanathan
  Kandeepan, and Ke~Wang.
\newblock Recurrent neural network (rnn) for delay-tolerant repetition-coded
  (rc) indoor optical wireless communication systems.
\newblock {\em Opt. Lett.}, 44(15):3745--3748, Aug 2019.

\bibitem{ML_ex3}
Nan Chi, Yiheng Zhao, Meng Shi, Peng Zou, and Xingyu Lu.
\newblock Gaussian kernel-aided deep neural network equalizer utilized in
  underwater pam8 visible light communication system.
\newblock {\em Opt. Express}, 26(20):26700--26712, Oct 2018.

\bibitem{ML_ex4}
Zhiquan Wan, Jianqiang Li, Liang Shu, Ming Luo, Xiang Li, Songnian Fu, and Kun
  Xu.
\newblock Nonlinear equalization based on pruned artificial neural networks for
  112-gb/s ssb-pam4 transmission over 80-km ssmf.
\newblock {\em Opt. Express}, 26(8):10631--10642, Apr 2018.

\bibitem{CNN_workload}
Teng Wang, Chao Wang, Xuehai Zhou, and Huaping Chen.
\newblock A survey of {FPGA} based deep learning accelerators: Challenges and
  opportunities.
\newblock {\em CoRR}, abs/1901.04988, 2019.

\bibitem{FPGAbeatsGPU}
Eriko Nurvitadhi, Ganesh Venkatesh, Jaewoong Sim, Debbie Marr, Randy Huang,
  Jason Ong Gee~Hock, Yeong~Tat Liew, Krishnan Srivatsan, Duncan Moss, Suchit
  Subhaschandra, and Guy Boudoukh.
\newblock Can fpgas beat gpus in accelerating next-generation deep neural
  networks?
\newblock In {\em Proceedings of the 2017 ACM/SIGDA International Symposium on
  Field-Programmable Gate Arrays}, FPGA '17, pages 5--14, New York, NY, USA,
  2017. ACM.

\bibitem{OptCom_gpu_ex1}
Sanjaya Lohani, Erin~M. Knutson, Matthew O'Donnell, Sean~D. Huver, and Ryan~T.
  Glasser.
\newblock On the use of deep neural networks in optical communications.
\newblock {\em Appl. Opt.}, 57(15):4180--4190, May 2018.

\bibitem{OptCom_gpu_ex2}
Jian Zhao, Yangyang Sun, Hongbo Zhu, Zheyuan Zhu, Jose~Enrique Antonio-Lopez,
  Rodrigo~Amezcua Correa, Sean Pang, and Axel Schülzgen.
\newblock {Deep-learning cell imaging through Anderson localizing optical
  fiber}.
\newblock {\em Advanced Photonics}, 1(6):1 -- 12, 2019.

\bibitem{CNN_ImageRecog}
Alex Krizhevsky, Ilya Sutskever, and Geoffrey Hinton.
\newblock Imagenet classification with deep convolutional neural networks.
\newblock {\em Neural Information Processing Systems}, 25, 01 2012.

\bibitem{FPGAbenefit}
Ahmad Shawahna, Sadiq~M. Sait, and Aiman El{-}Maleh.
\newblock Fpga-based accelerators of deep learning networks for learning and
  classification: {A} review.
\newblock {\em CoRR}, abs/1901.00121, 2019.

\bibitem{unrolled_opt}
Yufei Ma, Yu~Cao, Sarma Vrudhula, and Jae sun Seo.
\newblock Optimizing loop operation and dataflow in fpga acceleration of deep
  convolutional neural networks.
\newblock In {\em FPGA 2017 - Proceedings of the 2017 ACM/SIGDA International
  Symposium on Field-Programmable Gate Arrays}, pages 45--54. Association for
  Computing Machinery, Inc, 2 2017.

\bibitem{pipeline_opt}
{Huimin Li}, {Xitian Fan}, {Li Jiao}, {Wei Cao}, {Xuegong Zhou}, and {Lingli
  Wang}.
\newblock A high performance fpga-based accelerator for large-scale
  convolutional neural networks.
\newblock In {\em 2016 26th International Conference on Field Programmable
  Logic and Applications (FPL)}, pages 1--9, Aug 2016.

\bibitem{HW_leakyrelu}
K.~{Xu}, X.~{Wang}, and D.~{Wang}.
\newblock A scalable opencl-based fpga accelerator for yolov2.
\newblock In {\em 2019 IEEE 27th Annual International Symposium on
  Field-Programmable Custom Computing Machines (FCCM)}, pages 317--317, April
  2019.

\bibitem{RoF_CNN_BCNN}
Jeonghun Lee, Jiayuan He, Yitong Wang, Chengwei Fang, and Ke~Wang.
\newblock Experimental demonstration of millimeter-wave radio-over-fiber system
  with convolutional neural network (cnn) and binary convolutional neural
  network (bcnn).
\newblock {\em arXiv preprint arXiv:2001.02018}, 01 2020.

\bibitem{grad_vanishing}
Abien~Fred Agarap.
\newblock Deep learning using rectified linear units (relu).
\newblock {\em CoRR}, abs/1803.08375, 2018.

\bibitem{IEEE754_CNN}
Chen Zhang, Peng Li, Guangyu Sun, Yijin Guan, Bingjun Xiao, and Jason Cong.
\newblock Optimizing fpga-based accelerator design for deep convolutional
  neural networks.
\newblock In {\em Proceedings of the 2015 ACM/SIGDA International Symposium on
  Field-Programmable Gate Arrays}, FPGA ’15, page 161–170, New York, NY,
  USA, 2015. Association for Computing Machinery.

\bibitem{CortexA9_FPGA_ex1}
Li~Yang, Zhezhi He, and Deliang Fan.
\newblock A fully onchip binarized convolutional neural network fpga
  impelmentation with accurate inference.
\newblock In {\em Proceedings of the International Symposium on Low Power
  Electronics and Design}, ISLPED ’18, New York, NY, USA, 2018. Association
  for Computing Machinery.

\bibitem{Cortex_FPGA_ex2}
H.~{Yonekawa} and H.~{Nakahara}.
\newblock On-chip memory based binarized convolutional deep neural network
  applying batch normalization free technique on an fpga.
\newblock In {\em 2017 IEEE International Parallel and Distributed Processing
  Symposium Workshops (IPDPSW)}, pages 98--105, May 2017.

\end{thebibliography}





\end{document}